\documentclass[11pt]{article}
\usepackage{graphicx}
\input epsf
\usepackage{amsfonts, amsbsy, amssymb}

\begin{document}

\title{Quantum Creation of a Universe in Albrecht-Magueijo-Barrow model}
\author{A.V. Yurov*, V.A. Yurov**\\
\small *Department of Theoretical Physics, Russian
State University of Immanuel Kant,\\
\small        Aleksandra Nevskogo 14,
Kaliningrad 236041, Russia; e-mail {\em artyom\_yurov@mail.ru}\\
\small **Department of Mathematics, University of Missouri,\\
\small Columbia 65201, US; e-mail {\em valerian@math.mussouri.edu}
}



\maketitle

\maketitle
\begin{abstract}

In quantum cosmology the closed universe can spontaneously
nucleate out of the state with no classical space and time. The
semiclassical tunneling nucleation probability can be estimated as
$\emph{P}\sim\exp(-\alpha^2/\Lambda)$ where $\alpha$=const and
$\Lambda$ is the cosmological constant.

In classical cosmology with varying speed of light $c(t)$  it is
possible to solve the horizon problem, the flatness problem and
the $\Lambda$-problem if $c=sa^n$ with $s$=const and  $n<-2$. We
show that in VSL quantum cosmology with $n<-2$ the semiclassical
tunneling nucleation probability is
$\emph{P}\sim\exp(-\beta^2\Lambda^k)$ with $\beta$=const and
$k>0$. Thus, the semiclassical tunneling nucleation probability in
VSL quantum cosmology is very different from that in quantum
cosmology with $c$=const. In particular, it can be strongly
suppressed for large values of $\Lambda$. In addition, we propose
two instantons that describe the nucleation of closed universes in
VSL models. These solutions are akin to the Hawking-Turok
instanton in sense of $O(4)$ invariance but, unlike to it, are
both non-singular. Moreover, using those solutions we can obtain
the probability of nucleation which is suppressed for large value
of $\Lambda$ too.
\end{abstract}


\section{\label{sec:level1}Introduction}

One of the major requests concerning the quantum cosmology is a
reasonable specification of initial conditions in early universe,
that is in close vicinity of the Big Bang. The three wave
functions, describing the quantum cosmology has been proposed so
far: the Hartle-Hawking's \cite{4}, the Linde's \cite{5}, and the
so-called tunneling wave function \cite{6}. In the last case the
universe can tunnel through the potential barrier to the regime of
unbounded expansion. Following Vilenkin \cite{7} lets consider the
closed ($k=+1$) universe filled with radiation ($w=1/3$) and
$\Lambda$-term ($w=-1$). One of the Einstein's equations can be
written as a law of a conservation of the (mechanical) energy:
$P^2+U(a)=E$, where $P=-a\dot a$, $a(t)$ is the scale factor, the
''energy'' $E=$ const and the potential
$$
U(a)=c^2a^2\left(1-\frac{\Lambda a^2}{3}\right),
$$
where $c$ is the speed of light.
The maximum of the potential $U(a)$ is located at
$a_e=\sqrt{3/2\Lambda}$ where $U(a_e)=3c^2/(4\Lambda)$. The
tunneling probability in WKB approximation can be estimated as
\begin{equation}
\emph{P}\sim\exp\left(-\frac{2c^2}{8\pi G\hbar}\int_{a'_i}^{a_i}
da\sqrt{U(a)-E}\right), \label{1}
\end{equation}
where $a'_i<a_i$ are two turning points. The universe can start
from $a=0$ singularity, expand to a maximum radius $a'_i$ and then
tunnel through the potential barrier to the regime of unbounded
expansion with the semiclassical tunneling probability (\ref{1}).
Choosing $E=0$ one gets $a'_i=0$ and $a_i=\sqrt{3/\Lambda}$. The
integral in (\ref{1}) can be calculated. The result can be written
as
\begin{equation}
 \emph{P}\sim
\exp\left(-\frac{2c^3}{8\pi G\hbar\Lambda}\right). \label{2}
\end{equation}
For the probability to be of reasonable value, for example
$\emph{P}=1/{\rm e}\sim 0.368$, one has to put $\Lambda\sim
0.3\times 10^{65}\,\, {\rm cm}^{-2}$ (see (\ref{2})). In other
words, the $\Lambda$-term must be large. On the other side, the
universe once nucleated immediately begins a de Sitter
inflationary expansion. Therefore the tunneling wave function
results in inflation. And the $\Lambda$-term problem, which arises
in this approach is usually being gotten rid of via the anthropic
principle.
In this case we have two Lorentzian regions ($0<a<a'_i$, $a>a_i$)
and one Euclidean region ($a'_i<a<a_i$). The second turning point
$a=a_i$ corresponds to the beginning of our universe. If
$\Lambda=0$ then $U(a)$ has the form of {\bf parabola} and we get
only one Lorentzian region. In this case, the universe can start
at $a=0$, expand to a maximum radius and recollapse. If $E\to 0$,
the single Lorentzian region contracts to a single point, which
lies in agreement with the tunnelling nucleation probability:
$\emph{P}\to 0$ as $\Lambda\to 0$. However, as we'll show, in
quantum cosmological VSL models the situation can be opposite,
viz: the probability to find the finite universe short after it's
tunneling through the potential barrier is
$\emph{P}\sim\exp(-\beta(n)\Lambda^{\alpha(n)})$ with
$\alpha(n)>0$ and $\beta(n)>0$ when $n<-2$ or for $-1<n<-2/3$.
After the tunneling one gets the finite universe with "initial"
value of scale factor $a_i\sim\Lambda^{-1/2}$, so the probability
to find the universe with large value of $\Lambda$ and small value
of $a_i$ is strongly suppressed. The reason for this lies in the
behavior of potential $U(a)$, which, for the case $\Lambda\to 0$,
transforms into the {\bf hyperbola}, located under the abscissa
axis. As a result, such a universe can at $a\sim 0$ start the
regime of unbounded expansion. Therefore, we get the single
Lorentzian region that doesn't contract to a point at $E\to 0$.

This new property of VSL quantum cosmology will be discussed in
next Section but  new question arouse:  the geometric
interpretation of the quantum creation of a Universe with varying
speed of light. We know that universe can be spontaneously created
from nothing (when $c={\rm const}$) and this process can be
described with the aid of the instantons solutions possessing
$O(5)$ (if $V(\phi)$ has a stationary point at some nonzero value
$\phi=\phi_0={\rm const}$) or $O(4)$ (as Hawking-Turok instanton
\cite{HT}) invariance. So, what can be said about instantons in
the VSL models?

The whole plan of the paper looks as follows: in the next Section
we'll consider the simplest VSL model: model of
Albrecht-Magueijo-Barrow. Then we show that in framework of
tunneling approach to quantum cosmology with VSL the semiclassical
tunneling nucleation probability can be estimated as
$\emph{P}\sim\exp(-\beta^2\Lambda^k)$ with $\beta$=const and
$k>0$. All corresponding calculations will be done for the case of
the universe filled with radiation ($w=1/3$) and vacuum energy. In
the Section 3 we'll propose the {\bf non-singular} instanton
solutions possessing only $O(4)$ invariance (so the Euclidean
region is a deformed four sphere). These solutions can in fact
lead to inflation after the analytic continuation into the
Lorentzian region. We will discuss these results in Sec. 4.
\section {Albrecht-Magueijo-Barrow VSL model}
Lets start with the Friedmann and Raychaudhuri system of equations
with $k=+1$ (we assume the $G$=const):
\begin{equation}
\begin{array}{cc}
\displaystyle{\frac{\ddot a}{a}=-\frac{4\pi
G}{3}\left(\rho+\frac{3 p}{c^2}\right)+\frac{\Lambda
c^2}{3},\qquad \left(\frac{\dot a}{a}\right)^2=\frac{8\pi
G\rho}{3}-k\left(\frac{c}{a}\right)^2+\frac{\Lambda c^2}{3},}
\\
\\
\displaystyle{c=c_0\left(\frac{a}{a_0}\right)^n=sa^n,\qquad
p=wc^2\rho,}
\end{array}
\label{frid}
\end{equation}
where $a=a(t)$ is the expansion scale factor of the Friedmann
metric, $p$ is the fluid pressure, $\rho$ is the fluid density,
$k$ is the curvature parameter (we put $k=+1$), $\Lambda$ is the
cosmological constant, $c_0$ is some fixed value  of speed of
light which corresponds to a fixed value of scale factor  $a_0$.

Using (\ref{frid}) one gets
\begin{equation}
\dot\rho=-\frac{3\dot
a}{a}\left(\rho+\frac{p}{c^2}\right)+\frac{{\dot
c}c(3-a^2\Lambda)}{4\pi G a^2}. \label{rt}
\end{equation}
Choosing $w=1/3$ one can solve (\ref{rt}) to receive
\begin{equation}
\rho=\frac{M}{a^4}+\frac{3s^2na^{2(n-1)}}{8\pi G(
n+1)}-\frac{s^2n\Lambda a^{2n}}{8\pi G(n+2)}, \label{rho}
\end{equation}
where $M>0$ is a constant characterizing the amount of radiation.
It is clear from the (\ref{rho}) that the flatness problem can be
solved in a radiation-dominated early universe by an interval of
VSL evolution if $n < -1$, whereas the problem of $\Lambda$-term
can be solved only if $n<-2$.  The evolution equation for the
scale factor $a$ (the second equation in system (\ref{frid})) can
be written as
\begin{equation}
p^2+U(a)=E, \label{equation}
\end{equation}
where $p=-a\dot a$ is the momentum conjugate to $a$, $E=8\pi G
M/3$ and
\begin{equation}
U(a)=\frac{s^2a^{2n+2}}{n+1}-\frac{2s^2\Lambda a^{2n+4}}{3(n+2)}.
\label{U}
\end{equation}
The potential (\ref{U}) has one maximum at
$a=a_e=\sqrt{3/(2\Lambda)}$ such that
\begin{equation}
U_e\equiv
U(a_e)=\frac{s^23^{n+1}}{2^{n+1}\Lambda^{n+1}(n+1)(n+2)},
\label{Ue}
\end{equation}
so $U_e>0$ if (i) $n<-2$ or (ii) $n>-1$. The first case allows us
to solve the flatness and "Lambda" problems. Another benefit of
the model is a finite time region with accelerated expansion.
\subsection{\label{sec:level2}The semiclassical tunneling probability
in VSL models with $n<-2$: the case $E\ll U_e$} One can choose
$n=-2-m$ with $m>0$. Such a substitution gives us the potential
(\ref{U}) in the form
\begin{equation}
U_m(a)=\frac{s^2}{a^{2(m+1)}}\left(\frac{2\Lambda
a^2}{3m}-\frac{1}{m+1}\right). \label{Um}
\end{equation}

Since (\ref{equation}) is similar to equation of movement of the
particle of energy $E$ in the potential (\ref{Um}), the universe
in quantum cosmology can start at $a\sim 0$, expand to the maximum
radius $a'_i$ and then tunnel through the potential barrier to the
regime of unbounded expansion with "initial" value $a=a_i$. The
semiclassical tunneling probability can be estimated as
\begin{equation}
\emph{P}\sim\exp\left(-2\int_{a'_i}^{a_i} {\mid {\tilde p(a)}\mid}
da\right), \label{P}
\end{equation}
with
$$
{\mid {\tilde p(a)}\mid}=\frac{c^2(t)}{8\pi G\hbar}\mid
p(a)\mid,\qquad \mid p(a)\mid=\sqrt{U_m(a)-E},
$$
where $E\le U_e$. It is  convenient to write $E=U_e\sin^2\theta$,
with $0<\theta<\pi/2$.

For the case $E\ll U_e$ one can choose
\begin{equation}
\displaystyle{ a'_i\sim a_1=\sqrt{\frac{3m}{2(m+1)\Lambda}},\qquad
a_i\sim
\sqrt{\frac{3}{2\Lambda}}\left(\frac{\sqrt{m+1}}{\sin\theta}\right)^{1/m}
,} \label{ai-ai}
\end{equation}
 and evaluate the integral (\ref{P}) as
\begin{equation}
\displaystyle{
\emph{P}\sim\exp\left(\frac{-s^3\Lambda^{2+3m/2}I_m(\theta)}{4\pi
G\hbar}\right),} \label{P1}
\end{equation}
where
\begin{equation}
I_m(\theta)=\int_{z'_i(\theta)}^{z_i(\theta)} dz
z^{-5-3m}\sqrt{\frac{2z^2}{3m}-\frac{1}{m+1}}, \label{Im}
\end{equation}
with
$$
 z'_i(\theta)=\sqrt{\frac{3m}{2(m+1)}},\qquad
z_i(\theta)=\sqrt{1.5}\left(\frac{(m+1)^{1/2}}{\sin\theta}\right)^{1/m}.
$$
One can show that $I_m(\theta)>0$ at $0<\theta\ll 1$. Thus, it is
easy to see from (\ref{P1}) that the semiclassical tunneling
probability $\emph{P}\to 0$ for large values of $\Lambda>0$ and
$\emph{P}\to 1$ at $\Lambda\to 0$.

Note, that the case $c$=const can be obtained by substitution
$m=-2$ into the (\ref{P1}). Not surprisingly, this case will get
us the well known result $\emph{P}\sim\exp(-1/\Lambda)$ (see
\cite{7}).
\subsection{\label{sec:level3}The semiclassical tunneling probability with $n<-2$ and
$n>-1$}

In the case of general position the semiclassical tunneling
probability with $n=-2-m$ has the form
\begin{equation}
\emph{P}_m\sim \exp\left(-\frac{s^3\Lambda^{(3m+4)/2}}{4\pi G\hbar
3^{(m+1)/2}\sqrt{m(m+1)}}\int_{z'_i}^{z_i}
\frac{dz\sqrt{F_m(z,\theta)}}{z^{3m+5}}\right), \label{Pm}
\end{equation}
where
\begin{equation}
F_m(z,\theta)=-2^{m+1}\sin^2\theta\, z^{2(m+1)}+2\times 3^m (m+1)
z^2-m 3^{m+1}, \label{Fm}
\end{equation}
$z$ is dimensionless quantity and  $z'_i$, $z_i$ are the turning
points, i.e. two real positive solutions of the equation
$F_m(z,\theta)=0$ for the given $\theta$ ($F_m(z,\theta)=0$ does
have two such solutions at $0<\theta<\pi/2$).

If $m$ is the natural number then the expression (\ref{Pm}) has a
more simple form. For example
$$
\emph{P}_1\sim\exp\left(-\frac{s^3 \Lambda^{7/2}\sin\theta }{6\pi
G\hbar\sqrt{2}}\int_{z'_i}^{z_i}\frac{dz}{z^8}\sqrt{(z^2-{z'_i}^2)(z_i^2-z^2)}\right),
$$
with
$$
z'_i=\frac{\sqrt{3}}{2\cos(\theta/2)},\qquad
z'_i=\frac{\sqrt{3}}{2\sin(\theta/2)}.
$$
Similarly, $\emph{P}\sim\exp(-S)$, with
$$
S=\frac{s^3 \Lambda^5\sin\theta }{18\pi
G\hbar}\int_{z'_i}^{z_i}\frac{dz}{z^{11}}\sqrt{(z^2+z_1^2)(z^2-{z'_i}^2)(z_i^2-z^2)},
$$
where
$$
\begin{array}{cc}
z_1=\sqrt{\frac{3}{\sin\theta}\cos\left(\frac{\theta}{3}-\frac{\pi}{6}\right)},\qquad
z'_i=\sqrt{\frac{3}{\sin\theta}\sin\frac{\theta}{3}}, \qquad
z_i=\sqrt{\frac{3}{\sin\theta}\cos\left(\frac{\theta}{3}+\frac{\pi}{6}\right)},
\end{array}
$$
and so on.

Therefore the probability to obtain (via quantum tunneling through
the potential barrier) the universe in the regime of unbounded
expansion is strongly suppressed for large values of $\Lambda$ and
small values of the initial scale factor $a_i
=\sqrt{3}/(2\sin(\theta/2)\sqrt{\Lambda})$. In other words,
overwhelming majority of universes born via the quantum tunneling
through the potential barrier (\ref{U}) have large initial scale
factors and small values of $\Lambda$.


Now, let us consider the case (ii), when $n>-1$. The "quantum
potential" has the form
\begin{equation}
U(a)=s^2a^{2m}\left(\frac{1}{m}-\frac{2\Lambda
a^2}{3(m+1)}\right), \label{UU}
\end{equation}
where $m=n+1>0$. The points of intersection with the abscissa axis
$a$ are $a_0=0$ and $a_1=\sqrt{3(m+1)/2\Lambda m}$. Choosing $E=0$
in equation (\ref{equation}) and substituting (\ref{UU}) into the
(\ref{P}) we get
$$
\emph{P}\sim\exp\left(-\frac{s^3\Lambda^{(1-3m)/2}}{4\pi
G\hbar}\int_0^{z_1}
z^{2m-2}\sqrt{\frac{1}{m}-\frac{2z^2}{3(m+1)}}\,dz\right),
$$
with $z_1=\sqrt{3(m+1)/2m}$ (The starting value $z=0$ means that
the Universe tunneled from "nothing" to a closed universe of a
finite radius $a_1=z_1/\sqrt{\Lambda}$.). Thus, we have the same
effect as if $0<m<1/3$.


\subsection{\label{sec:level4}Peculiar cases with $n=-1$ and $n=-2$}

At last,  lets consider the cases of $n=-1$ and $n=-2$. The
formula (\ref{Pm}) is not valid in these cases ($m=-1$ and $m=0$)
so we shall consider these models separately.

If $n=-1$ ($m=-1$) then
$$
\rho=\frac{M}{a^4}+\frac{\Lambda s^2}{8\pi Ga^2}-\frac{3s^2}{4\pi
G a^4}\log\frac{a}{a_*},
$$
therefore
\begin{equation}
U(a)=s^2\left(2\log\left(\frac{a}{a_*}\right)-\frac{2a^2\Lambda}{3}+1\right),
\label{U-1}
\end{equation}
where $a_*$ is constant and $[a_*]$=cm. The potential (\ref{U-1})
has one maximum at $a=a_e=\sqrt{3/(2\Lambda)}$ such that
$U_e=U(a_e)=2s^2\log(a_e/a_*)$, so if $a_e>a_*$ then $U_e>0$. We
choose $a_*=\Lambda^{-1/2}$. This gives us $U_e=0.41 s^2>0$. For
the case $E\ll U_e$ the semiclassical tunneling nucleation
probability is
\begin{equation}
\begin{array}{cc}
\displaystyle{
\emph{P}_{_{-1}}\sim\exp\left(-\frac{s^3\sqrt{\Lambda}}{4\pi
G\hbar}\int_{z'_i}^{z_i}\frac{dz}{z^2}\sqrt{\log
z^2-\frac{2z^2}{3}+1}\right)\sim\exp\left(-\frac{s^3\sqrt{\Lambda}}{10\pi
G\hbar}\right),} \label{Pr1}
\end{array}
\end{equation}
where the turning points are $z'_i=0.721$, $z_i=1.812$. As we can
see from the (\ref{Pr1}), when $n=-1$ we receive the
aforementioned effect again.

If $n=-2$ ($m=0$) then
$$
\rho=\frac{M}{a^4}+\frac{s^2\Lambda}{2\pi
Ga^4}\log\left(\frac{a}{a_*}\right)+\frac{3s^2}{4\pi Ga^6}.
$$
We choose $a_*=1/(\alpha\sqrt{\Lambda})$, where $\alpha$ is a
dimensionless quantity. Thus
\begin{equation}
U(a)=-s^2\left(\frac{1}{a^2}+\frac{4\Lambda}{3}\log\left(\alpha
a\sqrt{\Lambda}\right)+\frac{\Lambda}{3}\right). \label{U-2}
\end{equation}
The maximum of potential (\ref{U-2}) is located at the same point
$a_e$ and
$$
U_e=-\frac{s^2\Lambda}{3}\left(3+\log\left(\frac{9\alpha^4}{4}\right)\right).
$$
Therefore, $U_e>0$ if $\alpha<2{\rm e}^{-3/4}/\sqrt{6}\sim 0.386$.
Choosing $\alpha=0.286$ and $E\ll U_e$ gets us the turning points
$z'_i\sim 0.77$ and $z_i\sim 2.391$.

At last, the semiclassical tunneling nucleation probability is
$$
\emph{P}_{0}\sim\exp\left(-\frac{s^3\Lambda^2}{4\pi
G\hbar}\int_{z'_i}^{z_i}\frac{dz}{z^4}\sqrt{-\frac{1}{z^2}-\frac{4}{3}\log(\alpha
z)-\frac{1}{3} }\right)\sim \exp\left(-\frac{0.084
s^3\Lambda^2}{\pi G\hbar}\right).
$$

\section{Instantons}
If we are going to describe the quantum nucleation of universe we
should find the instanton solutions, simply putted as a stationary
points of the Euclidean action. The instantons give a dominant
contribution to the Euclidean path integral, and that is the
reason of our interest in them.

First at all, lets consider the $O(4)$-invariant Euclidean
spacetime with the metric
\begin{equation}
ds^2=c^2(\tau)d\tau^2+a^2(\tau)\left(d\psi^2+\sin^2\psi
d\Omega_2^2\right). \label{metric}
\end{equation}
In the case $c={\rm const}$ one can construct the simple
instantons, which are the $O(5)$ invariant four-spheres. Then one
can introduce the scalar field $\phi$, whose (constant) value
$\phi=\phi_0$ is chosen as the one providing the extremum of
potential $V(\phi)$. The scale factor will be
$a(\tau)=H^{-1}\sin\,H\tau$ and after the analytic continuation
into the Lorentzian region one will get the de Sitter space or
inflation. Many other examples of non-singular and singular
instantons were presented in \cite{Linde-98}

Now, lets consider the VSL model with scalar field. The
corresponding Euclidean equations are:
\begin{equation}
\begin{array}{l}
\displaystyle{
\phi''+3\frac{a'}{a}\phi'=\frac{c^2V'}{\phi'}+\frac{c^5c'(\Lambda
a^2-3)}{4\pi G
a^2\phi'}+\frac{2\phi'c'}{c}-\frac{2cVc'}{\phi'},}\\
\\
\displaystyle{ \left(\frac{a'}{a}\right)^2=\frac{8\pi
G}{3c^4}\left(\frac{\phi'^2}{2}-c^2V\right)+\frac{c^2}{a^2}-\frac{\Lambda
c^2}{3}},
 \label{ins-1}
\end{array}
\end{equation}
where primes denote derivatives with respect to $\tau$.

At the next step we represent the potential $V$ in factorized form
\begin{equation}
V=F(a)U(\phi). \label{ins-2}
\end{equation}
Indeed, lets for example consider the power-low potential
$\sim\phi^k$. If the coupling $\lambda$ is dimensionless one then
we get
$$
V\sim \frac{\lambda}{\hbar} G^{k/2-2} c^{7-2k}\phi^k.
$$
Since $c=sa^n$ then in the simplest case we come to (\ref{ins-2}).

Let $\phi=\phi_0={\rm const}$ be solution of the (\ref{ins-1}).
(Note, that we don't require $\phi_0$ to be an extremum of
potential.) Using the first equation of system (\ref{ins-1}) and
(\ref{ins-2}) we get the equation for the $F(a)$,
\begin{equation}
\frac{dF(a)}{da}-\frac{2n}{a}F(a)=\frac{3ns^4}{4\pi GU_0
}a^{4n-3}-\frac{ns^4\Lambda}{4\pi G U_0}a^{4n-1} , \label{ins-3}
\end{equation}
where $U_0=U(\phi_0)={\rm const}$. The integration of the
(\ref{ins-3}) results in
\begin{equation}
F(a)=a^{2n}\left(C-\frac{3ns^4}{8\pi
G(1-n)U_0}a^{2(n-1)}-\frac{s^4\Lambda}{8\pi G U_0} a^{2n}\right),
\label{ins-4}
\end{equation}
where $C$ is the constant of integration and by assumption $n\neq
-1$ and $n\neq 0$. Substitution of (\ref{ins-4}) into the second
equation of the system (\ref{ins-1}) transforms it into the the
model of nonlinear oscillator, integration of which result in
\begin{equation}
 \frac{a'^2}{2}+u(a)=0,
\label{ins-5}
\end{equation}
where
\begin{equation}
u(a)=\frac{\omega^2 a^2}{2}-\frac{s^2 a^{2n}}{2(1-n)},
\label{ins-6}
\end{equation}
with $\omega^2=8\pi G U_0 C/(3s^2)$ and with the choice $C> 0$
made. We can see that for $c={\rm const}$ (i.e. $n=0$)
(\ref{ins-6}) turns out to be an equation of the harmonic
oscillator and we come to the well-known $O(5)$ solution (but in
this case $\phi_0$ must be the stationary point of $V$).

Equation (\ref{ins-5}) naturally describes the ''movement of a
classical particle'' with zero-point energy in mechanical
potential (\ref{ins-6}). Depending on value of $n$ this potential
can take one of four distinct forms (excluding the well-known
classical case $n=0$, which lies beyond the scoop of this
article).

\textbf{Case 1: $n<0$.} 
Here we have one Euclidean ($0\le a\le a_1$) and one Lorentzian
($a>a_1$) regions where
\begin{equation}
a_1=\left(\frac{s}{\omega\sqrt{1-n}}\right)^{1/(1-n)}.
\label{Intersection}
\end{equation}
On the bound between Euclidean and Lorentzian regions ($a=a_1$) we
have $a'=0$.

This mechanical potential is unbounded from below at $a\to 0$.
With this in mind, we'll have to ascertain that the Euclidean
action for our solution will stay finite. The gravitation action
has the form
$$
S_{{\rm grav}}=-\int d^4x \frac{c^3}{8\pi G}\sqrt{g} R.
$$
We are using the dimensionless variables $x^0=c_0\tau/a_0$,
$x^1=\psi$ and so on. Calculating $R$ we get
\begin{equation}
R=\frac{6}{c_0^2a^2}\left[c_0^2-\left(\frac{a_0}{a}\right)^{2n}\left((1-n)a'^2+aa''\right)\right],
\label{Ricci}
\end{equation}
so we do have the potential divergence at $a=0$. Multiplying
(\ref{Ricci}) on the $\sqrt{g}$ and $c^3$ and using the equation
of motion we get the expression:
\begin{equation}
R\sqrt{g} c^3\sim
6c_0\left((2-n)\omega^2\frac{a^{2n+3}}{a_0^{2n-1}}-\frac{nc_0^2}{1-n}\frac{a^{4n+1}}{a_0^{4n-1}}\right),
\label{Ricci-1}
\end{equation}
where the most dangerous multiplier factor is $a^{1+4n}$. But if
$-1/4 \le n < 0$ then the Euclidean action becomes finite and
therefore, we end up with the legitimate gravitation instanton. In
a similar manner, using (\ref{ins-2}) and (\ref{ins-4}) we get for
the scalar field (in dimensionless $x^{\mu}$):
$$
\sqrt{g}V_0\sim\frac{c_0a_0^{1-3n}}{8\pi G}(3\omega^2
a^{3(1+n)}+\frac{3nc_0^2a^{1+5n}}{(n-1)a_0^{2n}}-\frac{\Lambda
c_0^2a^{3+5n}}{a_0^{2n}})
$$
therefore the instanton exists for $n>-1/5$. This requirement is
stronger than the one for the gravitation instanton where $n>-1/4$
(see (\ref{Ricci-1})).

\textbf{Case 2. $0<n<1$.} Here the potential $u(a)$ suffers no
singularity at $a=0$, but $u(0)=0$. Also this potential has a
minimum at
$$
a_0=\left(\frac{s}{\omega}\sqrt{\frac{n}{1-n}}\right)^{1/(1-n)},
$$
and is equal to zero at (\ref{Intersection}), hence, once again
creating one Euclidean and one Lorentzian regions, separated by
(\ref{Intersection}).

\textbf{Case 3. $n=1$.} This case is somehow special, since for
such $n$ the solution of (\ref{ins-3}) shall be
$$
F(a)=a^{2n}\left(C-\frac{3s^4}{4\pi G U_0}
\ln{a}-\frac{s^4\Lambda}{8\pi G U_0} a^{2}\right),
$$
instead of (\ref{ins-4}), and hence, the equation of (\ref{ins-6})
shall be substituted by
\begin{equation}
u(a)=a^2 \left(\frac{\omega^2}{2}-s^2 \ln{a} \right).
\label{inss-1}
\end{equation}
It is easy to see that this function has two zeros (at $a_1=0$ and
$a_2=\exp(\frac{\omega^2}{2 S^2})$), is strictly positive on
interval ($a_1$, $a_2$) and strictly negative outside of it.
Therefore, this case doesn't allow an instanton.

\textbf{Case 4. $n>1$.} The potential $u(a)$ is strictly positive.
The instanton doesn't exist either.

Both of a newly founded solutions possess only $O(4)$ invariance
just like Hawking-Turok instanton (so the Euclidean region is a
deformed four sphere) but, unlike to it, they are all
non-singular. Note that if the value $a$ is sufficiently large
then one can neglect the second term in (\ref{ins-6}) (after the
analytic continuation into the Lorentzian region) therefore, as in
the case of  the usual $O(5)$ instanton, one can  get  the de
Sitter universe, i.e. the inflation.

The equation (\ref{ins-5}) has no terms with $\Lambda$. In other
words, the scale factor $a(\tau)$ doesn't depend on the value
$\Lambda$ (although being dependant on the $U_0$). Therefore, the
full Euclidean action $S_{_E}=S_{{\rm grav}}+S_{{\rm field}}$ has
the form,
$$
S_{_E}=S_0-\Lambda S_1,
$$
where $S_0$ and $S_1$ are both independent of the $\Lambda$.
Returning to what has been said in Introduction, there exist three
common ways to describe the quantum cosmology: the Hartle-Hawking
wave function $\exp(-S_{_E}/\hbar)$, the Linde wave function
$\exp(+S_{_E}/\hbar)$ and the tunneling wave function. In the
second Section we have been working with the tunneling wave
function. In case of instantons situation becomes slightly
different. If $S_1>0$ then (as a first, tree semiclassical
approximation) we should choose the Linde wave function, whereas
for the case $S_1<0$ the Hartle-Hawking wave function seems more
naturally.

In conclusion, we note that another choice of $C$ ($C<0$ and
$C=0$) eliminates all possible instantons.

\section{Discussion}

VSL models contain both some of the promising positive features
\cite{1}  and some shortcomings and unusual (unphysical?) features
as well \cite{3}. But, as we have shown, application of the VSL
principle to the quantum cosmology indeed results in amazing
previously unexpected observations. The first observation is that
the semiclassical tunneling nucleation probability in VSL quantum
cosmology is quit different from the one in quantum cosmology with
$c$=const. In the first case this probability  can be strongly
suppressed for large values of $\Lambda$ whereas in the second
case it is strongly suppressed for small values of $\Lambda$. This
is interesting, although we still can't say that VSL quantum
cosmology definitely results in solution of the $\Lambda$-mystery.
The problem here is the validity the WKB wave function. And what
is more, throughout the calculations we have been omitting all
preexponential factors (or one loop quantum correction) which can
be essential ones near the turning points. Another troublesome
question is the effective potentials in VSL models, being
unbounded from below at $a\to 0$. The naive way to solve this
problem is to use the Heisenberg uncertainty relation to find
those potentials with the ground state. However, this is just a
crude estimation. To describe the quantum nucleation of universe
we have to find the instanton solution which, being a stationary
point of the Euclidean action, gives the dominant contribution to
the Euclidean path integral. As we have seen, such solutions
indeed exist in VSL models. Those instantons are $O(4)$ invariant,
are non-singular, and provide an inflation as well. They describe
the quantum nucleation of universe from ''nothing'' and, what is
more, upon usage of these solutions we can obtain the probability
of a nucleation which is suppressed for large value of $\Lambda$
 using either Linde or Hartle-Hawking wave
function.

Note, that we can weaken the condition $n>-1/5$ to obtain a
singular instanton suffering the integrable singularity (i.e. such
that the instanton action will be finite) in the way of the
Hawking-Turok instanton. However, there exist some arguments
\cite{Vilenkin-98}, that such singularities, even being
integrable, still lead to serious problems with solutions.

In conclusion, we note that obtained instantons both have a free
parameter ($\omega^2$) so we are free to use the anthropic
approach to find the most probable values of $\Lambda$ too.

 \vspace{.2in} \noindent{\large {\bf Acknowledgements:}}
After finishing this work, we learned that T.Harko, H.Q.Lu,
M.K.Mak and K.S.Cheng \cite{Harko}, have independently considered
the VSL tunneling probability in both Vilenkin and Hartle-Hawking
approaches.  The interesting conclusion of their work is that at
zero scale factor the classical singularity is no longer isolated
from the Universe by the quantum potential but instead classical
evolution can start from arbitrarily small size. In contrast to
\cite{Harko}, we attract attention to the problem of
$\Lambda$-term and instantons in VSL quantum cosmology.

We'd like to thank Professor S. Odintsov for his encouraging
friendly support and Professor T. Harko for useful information
about the article \cite{Harko}.

$$
{}
$$
\bibliography{apssamp}

\vfill \eject

\end{document}